\documentstyle[aps,epsfig,floatfig,12pt]{revtex}

\begin{document}

\title{Heating of trapped ions from the quantum ground state\cite{gov}}
\author{
Q.\ A.\ Turchette,\cite{email} D.\ Kielpinski,  B.\ E.\ King,\cite{brian}
D.\ Leibfried,\cite{didi} D.\ M.\ Meekhof,
 C.\ J.\ Myatt,\cite{myatt}\\  M.\ A.\ Rowe, C.\ A.\ Sackett, C.\ S.\ Wood,\cite{wood} 
W.\ M.\ Itano, C.\ Monroe, and D.\ J.\ Wineland
}
\address{
Time and Frequency Division, 
National Institute of Standards and Technology,
Boulder, CO 80303
}
\date{\today}
\maketitle

\begin{abstract}
We have investigated motional heating of laser-cooled $^9$Be$^+$ ions held in
radio-frequency (Paul) traps.
We have measured heating rates in a variety of traps with different
geometries, electrode materials, and 
characteristic sizes. The results show that heating 
is due to electric-field noise
from the trap electrodes which exerts a stochastic fluctuating force on the ion.
The scaling of the heating rate with trap size is much stronger than that
expected from a spatially uniform noise source on the electrodes
(such as Johnson noise from external circuits),
indicating that a microscopic uncorrelated noise source on the electrodes
(such as fluctuating patch-potential fields) is a more likely candidate
for the source of heating.
\end{abstract}


\section{Introduction}

Cold trapped ions have been proposed as a physical implementation
for quantum computation (QC) \cite{Cirac95}, and experiments on one 
\cite{Monroe95,Monroe95a,Meekhof96,Roos99} 
and two \cite{King98,Turchette98} ions have
demonstrated proof of the principle. Work is currently underway
to extend these results. In ion trap
QC, ion-laser interactions prepare, manipulate and entangle atomic states
in ways dependent on the quantum motional
state of the ions. A limiting factor
in the fidelity of an operation is uncontrolled heating of the motion
during manipulations. Heating leads
to decoherence of the quantum superposition states involved
in the computation \cite{Wineland98,Myatt99}, and
can ultimately limit the number of elementary 
gate operations which can be strung together. 
Speculations have been made about the mechanisms 
that lead to heating 
\cite{Wineland98,Diedrich89,Wineland98a,Lamoreaux97,James98,Henkel99}, 
but measurements are scarce since the necessary
sensitivity  can be achieved only through laser cooling to near the
ground state of motion. Additionally, systematic studies of the
dependence of heating rate on various trap properties are difficult,
since often this requires the construction and operation of
an entirely new trap apparatus which may have  different values
of properties not under study.  Indeed, the data presented here
pose several interpretational difficulties for this reason.

Heating of a single trapped ion (or the center-of-mass motion of
a collection of trapped ions) occurs when noisy 
electric fields at the position of the ion 
couple to its charge, giving rise to fluctuating forces. If the
spectrum of fluctuations overlaps 
the trap secular motion frequency 
or its micro-motion sidebands, 
the fluctuating forces
can impart significant energy to the secular motion of the ion.
Here, we express the heating rate as the average number of quanta of 
energy gained by the secular motion in a given time.
There are several candidates worth considering for sources of the noisy fields
which give rise to heating.  Some of these are \cite{Wineland98}:
 Johnson noise from the resistance
in the trap electrodes or external circuitry 
(the manifestation of thermal electronic
noise or black body radiation consistent with the boundary
conditions imposed by the trap electrode structure),
fluctuating patch-potentials (due, for example, to randomly oriented 
domains at the surface of the electrodes or adsorbed materials on the 
electrodes), ambient electric fields from
injected electronic noise, fields generated by fluctuating currents 
such as electron currents from field-emitter points on the trap electrodes,
and collisions with background atoms. 
Only the first two
mechanisms will be considered here since the remaining mechanisms
(and others) are
unlikely contributers \cite{Wineland98} or can be eliminated by
comparing the measured heating rates of the center-of-mass and
differential modes of two ions \cite{King98}.
As will be shown below, 
the Johnson-noise and patch-potential mechanisms 
give rise to heating rates which scale differently 
with the distance between the ion and the 
trap electrodes. 

\section{Two models for sources of heating}

\subsection{Preliminaries}

The heating rate caused by a fluctuating uniform field can be derived as in
Savard, {\it et al.} \cite{Savard97} and agrees with a classical calculation
\cite{Wineland98}.  The Hamiltonian 
for a particle of charge
$q$ and mass $m$ trapped in an harmonic well
subject to a fluctuating, uniform (non-gradient) electric field drive $\epsilon (t)$ is
\begin{equation}
H(t) =  H_0 - q \epsilon(t) x,
\end{equation} 
where $H_0 = p^2/2m + m\omega_m^2x^2$ is the usual, 
stationary harmonic oscillator Hamiltonian with 
trap frequency $\omega_m$.
From first order perturbation theory, the rate of transition
 from the ground state of the well
($|n=0\rangle$) to the first excited state ($|n=1\rangle$) is \cite{Savard97}
\begin{equation}
\Gamma_{0 \rightarrow 1} = \frac{1}{\hbar^2} \int_{-\infty}^{\infty}d \tau e^{i\omega_m\tau}
\langle \epsilon(t)\epsilon(t+\tau)\rangle |\langle 0 |q x | 1 \rangle |^2.
\end{equation}
Evaluating the motional matrix element gives
\begin{equation}
\Gamma_{0 \rightarrow 1} = \frac{q^2 }{m\hbar\omega_m} S_E(\omega_m)
\end{equation}
where $S_E(\omega) \equiv \int_{-\infty}^{\infty} d \tau e^{i\omega_m\tau}
\langle \epsilon(t)\epsilon(t+\tau)\rangle$
is the spectral density of electric-field fluctuations.

For an  
ion trapped by a combination of
(assumed noiseless) static and inhomogeneous rf fields (Paul trap)
the heating rate can be generalized to \cite{Wineland98a}:
\begin{equation}
\dot {\bar n} = \frac{q^2 }{4m\hbar\omega_m}\left(S_E(\omega_m) + 
\frac{\omega_m^2}{2\Omega_T^2}S_E(\Omega_T \pm \omega_m) \right),
\label{eq:heatrate}
\end{equation}
where $\dot {\bar n}$ is the rate of change of the average
thermal occupation number,
$\omega_m$ is now the secular frequency of the mode of motion under 
consideration and $\Omega_T$ is the trap rf-drive frequency.  
The second term 
on the rhs of Eq.\ \ref{eq:heatrate} is due to a cross-coupling
between the rf and noise fields; it
 will not be present for the axial motion of
a linear trap which is confined only by static fields.  Even for
motion confined by rf pondermotive forces,
 this second term
will be negligible in the absence of spurious resonances in $S_E(\omega)$ or increasing
$S_E(\omega)$ (since $\omega_m^2/\Omega_T^2 \sim 10^{-4}$) 
and is neglected in what follows \cite{Wineland98}.


We differentiate two sources of the noise that gives rise to heating.  
The first is thermal electronic noise in the imperfectly conducting
trap electrodes and elsewhere in the trap circuitry. Though this source
of noise is ultimately microscopic in origin, for our purposes here
it can be treated adequately by use of lumped circuit models. 
Thermal noise has been considered in the context of ion-trap heating
in several places \cite{Wineland98,Wineland98a,Lamoreaux97,James98,Henkel99}.
The second source of noise considered here is due to 
``microscopic'' regions of material (small compared to the size of the
trap electrodes)
with fluctuating, discontinuous potentials established, for example, 
at the interface of different materials or crystalline domains.  
We call this patch-potential noise, and
its microscopic origin leads to
manifestly different heating behaviour from that for the thermal electronic noise case. Static
patch-potentials are a well-known phenomenon, but little is known
about the high-frequency (MHz) fluctuating patches which are required to
account for our observed heating rates \cite{Camp91,Kleint63,Timm66}.

\subsection{Thermal electronic noise}

Heating rates in the case of thermal electronic noise (Johnson noise)  
can be obtained simply 
through the use of lumped-circuit models, which are 
justified by the fact that the wavelength
of the relevant fields (at typical trap secular or drive frequencies) 
is significantly
larger than the size of the trap electrodes.  Such an analysis 
has been carried out 
elsewhere \cite{Wineland98,Wineland98a,Wineland75}, and only the major 
results will be quoted here.
Resistances in the trap electrodes and connecting circuits 
give rise to an electric field noise spectral density
$S_E(\omega) = 4k_BTR(\omega)/d^2$ where $d$ is the 
characteristic distance from
the trap electrodes to the ion, $T$ is the temperature 
(near room temperature for all of our experiments), $k_B$ is 
Boltzmann's constant, and $R(\omega)$ is the effective 
(lumped-circuit) resistance between trap electrodes. 
The heating rate is given by
\begin{equation}
\dot{\bar n}_{\rm R} = \frac{q^2 k_BTR(\omega_m)}{m\hbar\omega_m d^2}. 
\label{eq:heatrateR}
\end{equation}
A numerical estimate of the heating rate for typical trap parameters
gives $0.1$/s $< \dot{\bar{n}}_{\rm R} < 1$/s \cite{Wineland98,Wineland98a},
which is significantly slower than our observed rates.
As a final note,
the lumped circuit approach is convenient, but not necessary.
In the Appendix, we present a microscopic
model that is valid for arbitrary ion-electrode 
distances and reproduces Eq.\ \ref{eq:heatrateR} for all the traps
considered here (and for all realistic traps where $d \gg \delta$,
where $\delta$ is the skin depth of the electrode material at the
trap secular frequency).

\subsection{Fluctuating patch-potential noise}

To derive the heating rate for the case of microscopic patch-potentials
we use the following approximate model.  We assume that the trap
electrodes form a spherical conducting shell of radius $a$ around the ion. 
Each of the patches is a disc on the inner surface 
of the sphere with radius $r_p \ll a$
and electric potential noise $V_p(\omega)$.  Alternatively,
each patch is assumed to have power noise spectral density $S_V(\omega)$.
The electric field noise at the ion due to a single patch
is $E_p(\omega) = -3V_p(\omega)r_p^2/4a^3$
in the direction of the patch. 
There are $N \approx 4Ca^2/r_p^2$ 
such patches distributed over
the sphere with coverage $C \leq 1$.  Averaging over a random 
distribution of patches on the sphere, we find that 
the power spectral density of the electric field at the ion
(along a single direction) is 
\begin{equation}
S_E(\omega) = N 
\left( \frac{\partial E_p(\omega)}{\partial V_p(\omega)} \right) ^2
S_V(\omega) = \frac{3CS_V(\omega)r_p^2}{4a^4}.
\end{equation}
This gives a heating rate
\begin{equation}
\dot{\bar{n}}_{\rm P} = \frac{3q^2 Cr_p^2 S_V(\omega_m)}{16m\hbar\omega_m d^4 },
\label{eq:heatrateP}
\end{equation}
in which the association $d \sim a$ is made.
Note the difference in scaling with electrode size between 
Eqs. \ref{eq:heatrateR}
and \ref{eq:heatrateP}. The thermal electronic noise model gives a scaling
$\dot{\bar{n}}_{\rm R} \propto d^{-2}$, while the patch-potential
model gives $\dot{\bar{n}}_{\rm P} \propto d^{-4}$.  In fact, a $d^{-4}$
dependence also arises from a random distribution of fluctuating
charges or dipoles.

\section{Measurements}

\subsection{Measuring the heating rate}
 
To determine the heating rate, we first cool the ion to 
near the ground state.  In sufficiently strong traps, 
this is achieved simply
by laser cooling with light red-detuned from
a fast cycling transition ($\gamma \approx \omega_m$,
where $\gamma$ is the radiative linewidth of the upper state)    
propagating in
a direction such that its $k$-vector has a component along the 
direction of the mode of interest.
In weaker traps, additional sideband Raman cooling is 
utilized to cool to the ground
state \cite{Monroe95a}.  
Typical starting values of $\bar n$, 
the average number of thermal phonons
in the mode of interest, are between $0$ and $2$.  

After cooling and optically pumping the ion to its internal ground
state (denoted $|\downarrow\rangle$), we drive Raman 
transitions between atomic and motional levels \cite{Monroe95,Monroe95a,Meekhof96}. 
Tuning the Raman difference frequency $\Delta \omega$ to the $k^{th}$ 
motional blue sideband (bsb) at $\Delta \omega = \omega_0 + k\omega_m$  
drives the transition $\vert\downarrow\rangle\vert n\rangle
\leftrightarrow \vert\uparrow\rangle\vert n+k\rangle$ 
where $\vert \uparrow \rangle , \vert\downarrow\rangle$
refer to the internal (spin) states of the atom that are
separated by $\omega_0$. The $k^{th}$ red sideband (rsb)
at $\Delta \omega = \omega_0 - k\omega_m$ drives $\vert\downarrow\rangle\vert n\rangle
\leftrightarrow \vert\uparrow\rangle\vert n-k\rangle$.
The measurement utilizes asymmetry in the strengths of
the red and blue motional sidebands to extract $\bar n$.
The strengths of the sidebands are defined as the probability of making
a transition $|\downarrow\rangle \leftrightarrow |\uparrow\rangle$,
which depends on the occupation number of the motional levels.
The strengths are probed by a Raman pulse of duration $t$ tuned to either
$k^{th}$ sideband.
The probability $P_\downarrow$ of remaining in $|\downarrow\rangle$ after probing
 is measured and the strengths 
$I_k^{\rm rsb} = 1-P_{\downarrow,{\rm rsb}}$ and
$I_k^{\rm bsb} = 1-P_{\downarrow,{\rm bsb}}$ are extracted.
For thermal motional states,
 the strengths of the red and blue sidebands are related by \cite{Monroe95a,Wineland98}
\begin{eqnarray}
I_k^{\rm rsb} & = & \sum_{m=k}^\infty P_m  \sin^2 \Omega_{m,m-k} t \\
& = &  \left( \frac{\bar n}{1+\bar n} \right)^k \sum_{m=0}^\infty P_m  \sin^2 \Omega_{m+k,m} t \\
& = & \left( \frac{\bar n}{1+\bar n} \right)^k I_k^{\rm bsb}
\end{eqnarray}
where $\Omega_{m+k,m} = \Omega_{m,m+k}$ is the Rabi frequency of the
$k^{th}$ sideband between levels $m$ and $m+k$, and $P_m = {\bar n}^m/(1+\bar n)^{m+1}$
is the probability of the $m^{th}$ level being occupied for a thermal distribution of
mean number $\bar n$.
The ratio of the sidebands $R_k \equiv I_k^{\rm rsb}/I_k^{\rm bsb}$ 
is independent of drive time $t$ and immediately gives the
mean occupation number $\bar n$,
\begin{equation}
\bar n = \frac{(R_k)^{1/k}}{1-(R_k)^{1/k}},
\label{eq:nbar}
\end{equation}
which is valid even if the Lamb-Dicke criterion is not satisfied.
In principle, $k$ should be chosen to be the
positive integer nearest to $\bar n$ in order to maximize sensitivity.
In practice we use $k = ${1, 2, or 3} in most cases.
 Note that Eq.\ \ref{eq:nbar} is valid only for
thermal states; this is adequate since Doppler cooling leaves 
the motion in a thermal state
\cite{Meekhof96,Stenholm86}, as does any cooling to near the ground state.

\begin{figure}
\epsfig{file=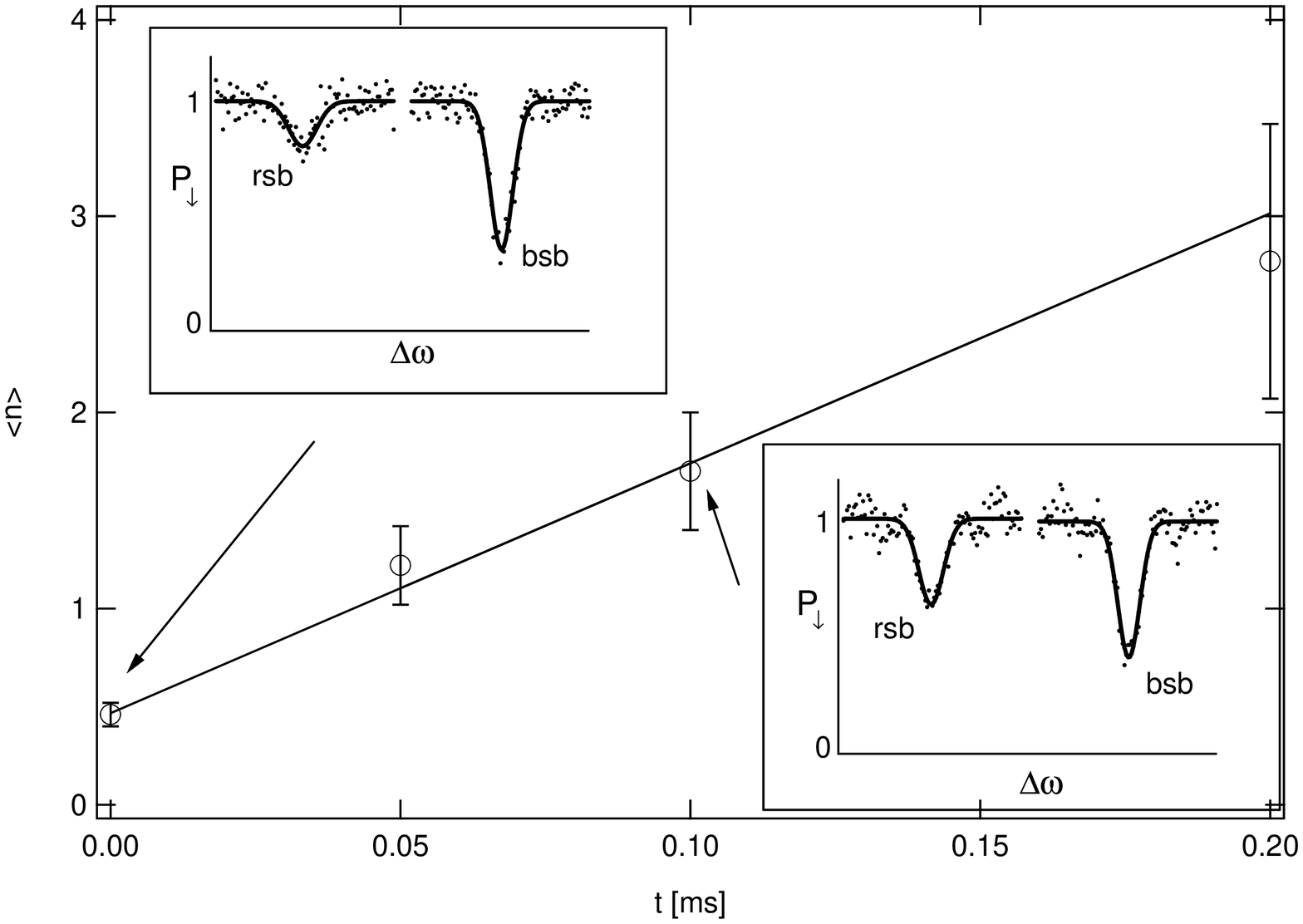,width=\textwidth}
\caption{An example of heating rate data.  
The main graph shows $\bar n \; (\equiv \langle n \rangle)$ vs $t$, 
the delay between cooling and probing.  The insets show Raman 
spectra from which
$\bar n$ is extracted, according to Eq.\ \protect\ref{eq:nbar}.
For the insets, $P_\downarrow$ is the probability that the ion remains
in the $|\downarrow\rangle$ state after application of a 
Raman probe of fixed duration
with difference frequency $\Delta\omega$; 
rsb: red motional sideband, bsb: blue motional sideband. 
The sidebands shown are the 1st sidebands. The data are
for trap 5 from Table \protect\ref{table:traps}
at 5 MHz secular frequency and $\dot {\bar n} = 12 \pm 2$/ms.
}
\label{fig:dndt}
\end{figure}

In order to determine the heating rate $\dot{\bar n}$,
delays with no laser interaction are added between the 
cooling cycle and the probing cycle. An example
of a data set at a fixed trap secular frequency is 
shown in Fig.\ \ref{fig:dndt}.
The error bars are determined as follows:  The raw data 
of Raman scans over the sidebands
(such as is those shown in the insets of
Fig.\ \ref{fig:dndt}) are fit to Gaussians, from which
the depths of the sidebands are extracted, with error on the parameter 
estimate calculated assuming normal distribution of the data. 
The errors from the rsb and bsb strengths
are propagated through Equation \ref{eq:nbar} for an error on $\bar n$.
The error bars shown are one-sigma, 
and include only statistical factors.
These errors are incorporated in the linear 
regression to extract $\dot{\bar{n}}$ with
appropriate error. 
Many such data sets are taken for various types 
of traps and at different secular frequencies.

\subsection{The traps}

\begin{figure}
\epsfig{file=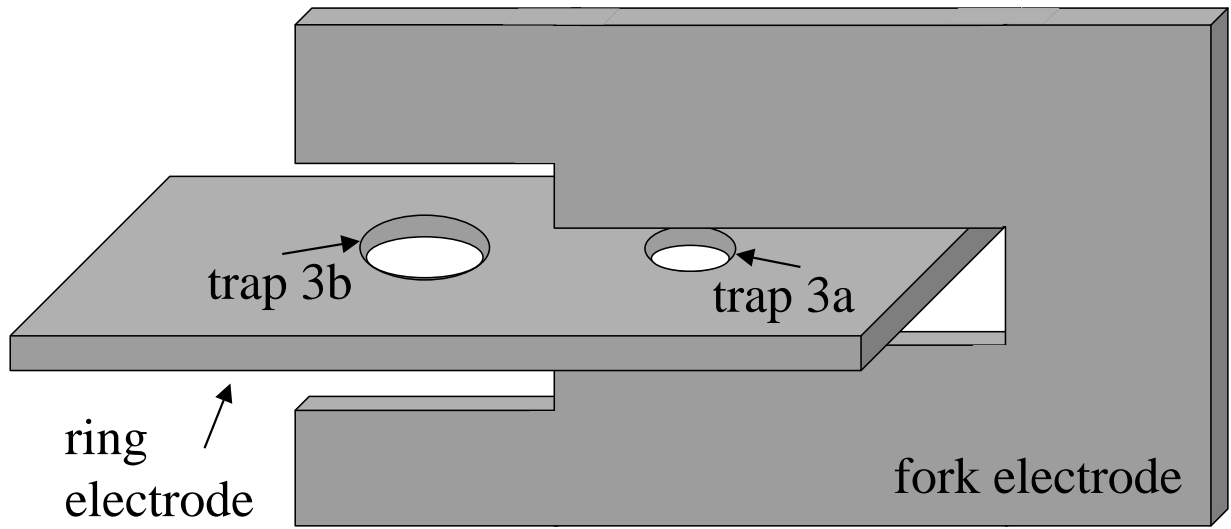,width=\textwidth}
\caption{Schematic diagram of the electrodes of  
trap 3 (from Table \protect\ref{table:traps}).
The distance between traps 3a and 3b is 1.7 mm.
Not to scale.
}
\label{fig:dtrap}
\end{figure}

\begin{figure}
\epsfig{file=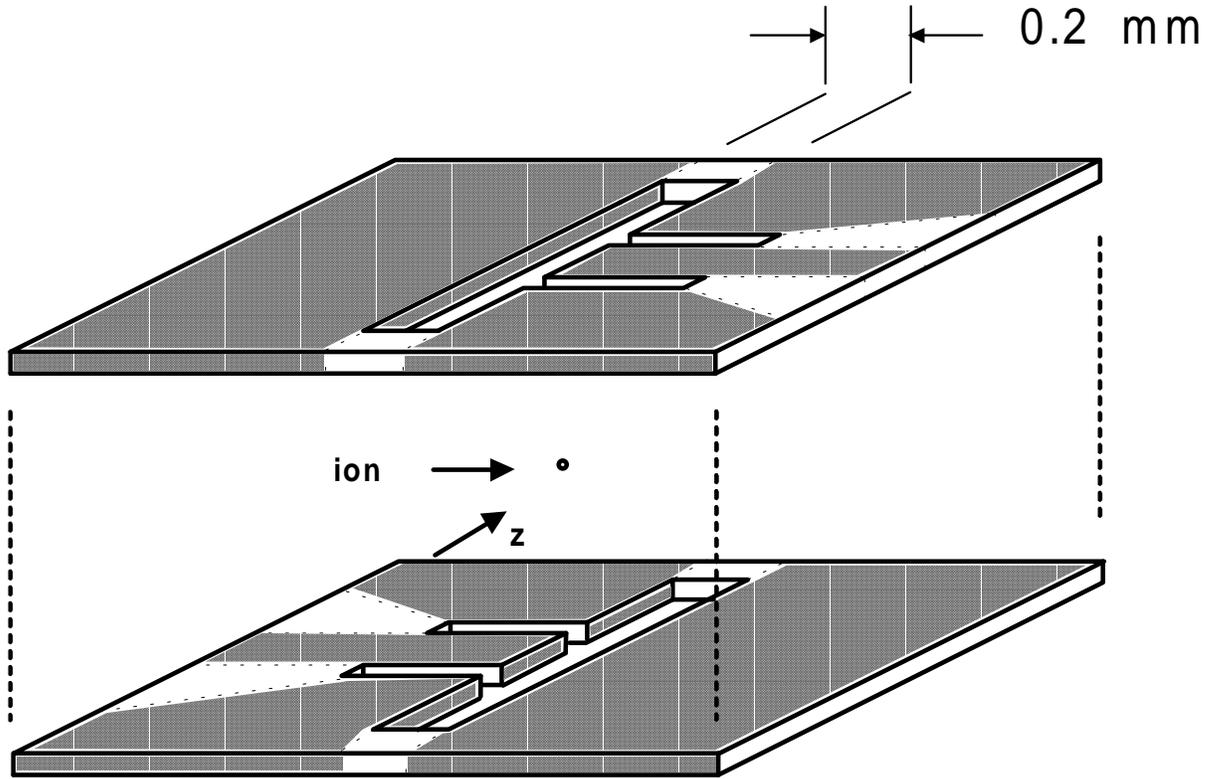,width=\textwidth}
\caption{Schematic diagram of the electrodes of the linear traps 
(traps 4, 5 and 6 from Table \protect\ref{table:traps}).
The traps are formed by evaporating gold 
(approximately 0.75 $\mu$m thick) on an alumina substrate.
The outer segmented electrodes are the endcaps, while the
long unbroken electrodes carry rf. The axial direction (labeled $z$) 
is parallel to the rf electrode. The two separate trap
wafers are spaced by 200 $\mu$m for traps 4 and 5 and 280 $\mu$m
for trap 6 (spacers not shown).
Schematic diagram  not to scale.
}
\label{fig:lintrap}
\end{figure}

The measurements of heating rate  in this paper extend over a five year period
and utilize six different traps. 
The traps are summarized in Table \ref{table:traps}.
The traps are described in the references listed; here only
a brief discussion is included.  
The ``ring'' traps are approximate quadrupole configurations consisting
 of a flat electrode ($125 \; \mu$m thick) with a hole
drilled through it (the ring) and an independent ``fork'' 
electrode ($100 \; \mu$m thick) that forms endcaps on either
side of the electrode, centered with the hole, similar to the trap
shown in Fig.\ \ref{fig:dtrap}.
In trap 1, the ring and endcap electrodes are at the same average potential; 
in traps 2 and 3 a static bias field could be added between the fork 
and ring to change the distribution
of binding strengths along the three principle axes of the trap. 
The size of these traps is stated as the hole radius, 
with the endcaps spacing approximately 70\% of the hole diameter.
For the elliptical ring trap (trap 2) the stated size is the radius
along the minor axis and the aspect ratio is 3:2; the
fork tines are parallel to the major axis of the ellipse.
Traps 3a and 3b were drilled into a single flat electrode
with a single graded fork electrode (see Fig.\ \ref{fig:dtrap}).
The rings are circular and the size stated is the radius.
This was the trap used for the size-scaling measurements.
The heating in all of the ring traps was measured in
a direction in the plane of the ring electrode,
 parallel to the tines of the fork electrode.
Traps 4, 5 and 6 are similar linear traps
with geometry indicated in Fig.\ \ref{fig:lintrap}.
Trap 6 was made slightly larger than traps 4 and 5
by increasing the space between the two electrode wafers.
Heating was measured along the axial direction, which has only a static
confining potential.
The size quoted in Table \ref{table:traps} for the linear traps is 
the distance between the ion and the nearest electrode.
All traps are mounted at the end of a coaxial $\lambda$/4 resonator 
for rf voltage buildup \cite{Jefferts95}. 
Typical resonator quality factors are
around 500 and rf voltage at the open end is approximately 
500 V with a few watts of input power.
In all traps except for trap 3a and 3b 
the resonator is inside the vacuum chamber with the
trap. In traps 3a and 3b, the resonator is outside the chamber, with the
high-voltage rf applied to the trap through a standard vacuum feedthrough.

Since we believe that surface effects are an important factor in heating,
we cleaned the electrode surfaces before using a trap.  When trap electrodes
were recycled, they were first cleaned with HCl in order to remove the Be
coating deposited by the atomic source.
For the molybdenum traps an electro-polish in phosphoric acid was then used.
For the beryllium electrodes electro-polishing in a variety of
acids was ineffective, so abrasive polishing was used. 
Finally, the traps were rinsed in distilled water followed by methanol.
The gold electrodes of the linear traps were cleaned with solvents
after being evaporatively deposited on their alumina substrates. 
The time of exposure of clean trap electrodes to the atmosphere before
the vacuum chamber was evacuated  was typically less than one day. 
The traps were then vacuum baked at $\sim 350^\circ$ C for approximately
three days.

\subsection{Data}

\begin{figure}
\epsfig{file=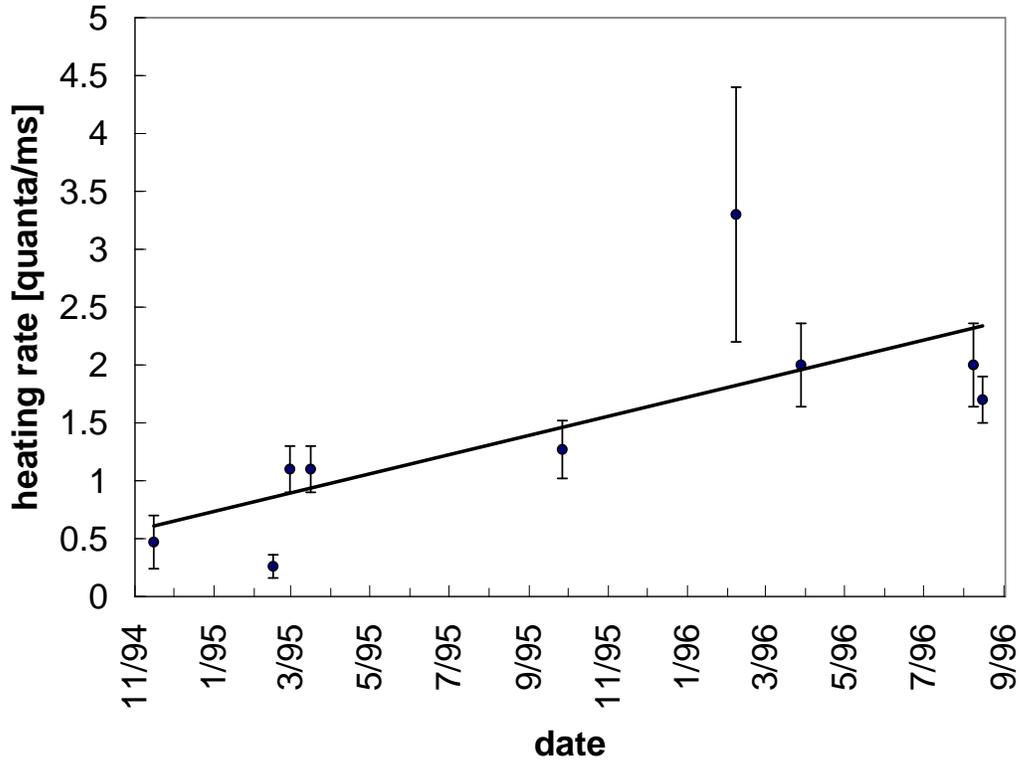,width=\textwidth}
\caption{Heating rate in ring trap 1 {\it vs.} time. The 
secular frequency for all measurements was $\sim 11$ MHz. The solid
line shows a trend which does not account for the weights of the data
points.
}
\label{fig:molyt}
\end{figure}

Our longest-term heating measurements were made on trap 1. 
In Figure \ref{fig:molyt}
we plot the heating rate as a function of date of data acquisition 
for a fixed trap frequency ($11$ MHz). The heating rate
is on the order of 1 quantum per millisecond with 
a basic trend upwards of $\sim 1$ quantum per millisecond per year.  
Over this time
the electrodes were coated with Be from the source ovens,
 but beyond this, nothing
was changed in the vacuum envelope, which was closed for this entire period
of time. The cause of the increase in heating rate is unknown, but may be
related to increased Be deposition on the electrodes. Be plating on the 
trap electrodes could be a source of patch-potential noise.

\begin{figure}
\epsfig{file=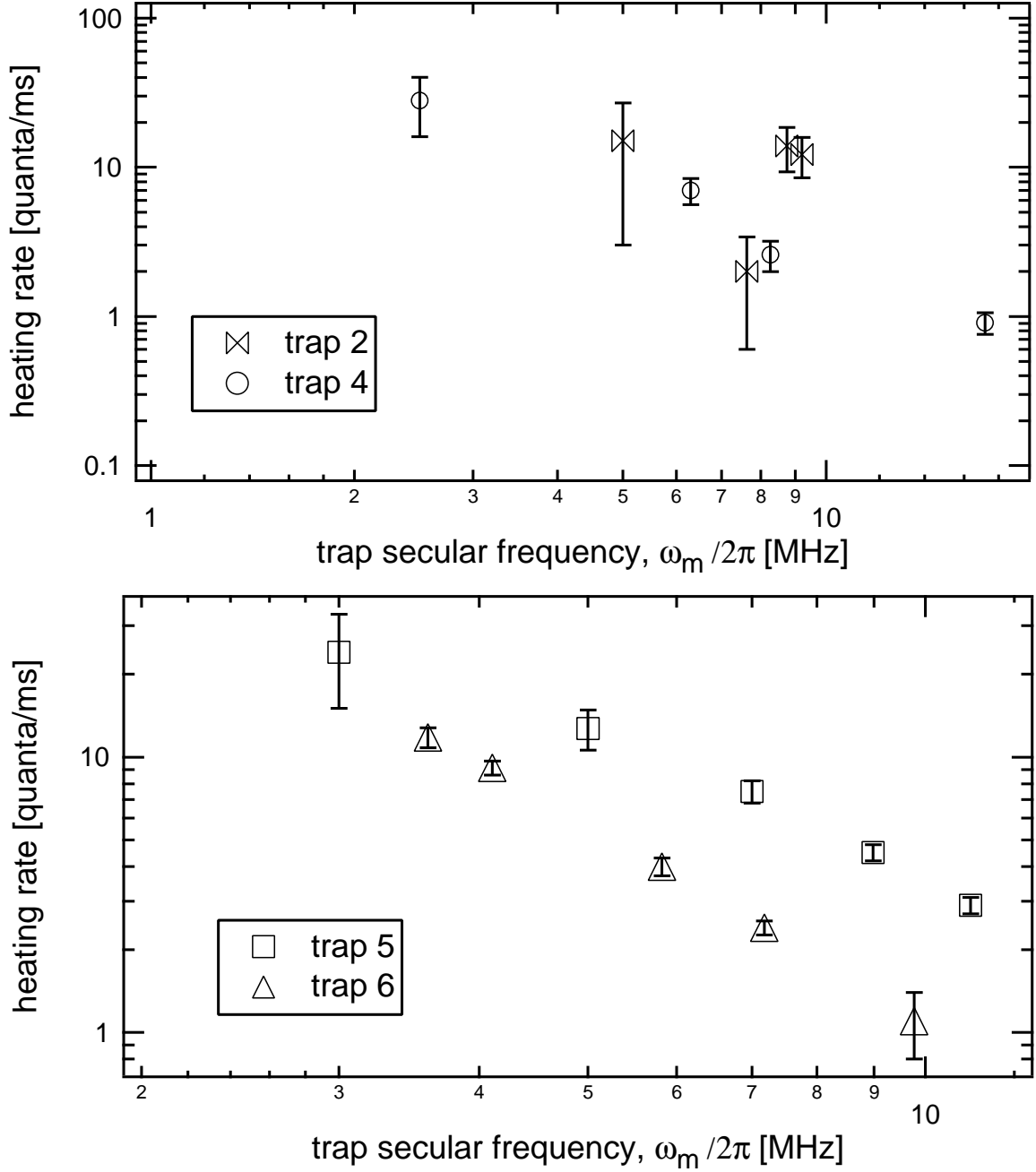,width=\textwidth}
\caption{Heating rates {\it vs.} trap secular frequency
in a) the elliptical ring trap 2 
and in the micro linear-trap 4 and in b) linear traps 5 and 6.
The only intended difference between traps 5 and 6 is the size. 
 In all four data sets, the secular frequency was 
varied by changing a static potential only.
}
\label{fig:didilin}
\end{figure}

Figure \ref{fig:didilin} shows heating rates in the linear traps
(trap 4, 5 and 6) and the elliptical ring trap (trap 2)
as a function of trap secular frequency. 
The frequency dependence of the heating rate is expected to 
scale as $S_E(\omega_m)/\omega_m$ (Eq.\ \ref{eq:heatrate}).
For example, a trap electrode with a flat noise spectrum
($S_E(\omega) =$ constant) will have a heating rate 
that scales as $\omega_m^{-1}$.  The actual spectrum
of fluctuations is impossible to know {\it a priori}, but
in principle the data can be used to extract a spectrum over a 
limited frequency range given the model leading to Eq.\ \ref{eq:heatrate}.
For the three linear traps, the heating rate data are most
consistent with a $\omega_m^{-2}$ scaling, implying  $S_E \propto \omega^{-1}$.
This does not
greatly assist in identifying a physical mechanism for the heating.  
For example, pure Johnson noise will 
have a flat spectrum, low-pass-filtered Johnson noise will have a spectrum
that decreases with increasing frequency, and
the spectrum of fluctuations in the patch-potential case is
entirely unknown.  In addition to the theoretical ambiguity, there is
evidence in other data  sets of different frequency scalings (though they are
always power-law scalings).  
This measurement certainly cannot be used to pinpoint a heating mechanism; it is presented here
only for completeness.

The data of Figure \ref{fig:didilin}b provide a first indication of the
scaling of the heating rate with trap size.  Trap 6 is about 1.3 times larger
than trap 5, while its heating rate (at 10 MHz) was a factor of 3 slower.
This indicates that the dependence of heating rate on trap size
is stronger than $d^{-2}$, but is consistent with $d^{-4}$.  Of course, this
comparison is to be taken with some caution, since these are two separate traps
measured several weeks apart, and therefore likely had different
microscopic electrode environments. However, a comparison is warranted since
the traps were identical apart from their sizes.
In particular, all the associated electronics was the same and the rf drive voltage
was very nearly the same.  In fact, the rf voltage was slightly larger for the measurements on trap 6,
which showed the lower heating rate. This is important to note because we observe
a slight dependence of the heating rate on the applied rf trapping voltage. Though
we have only a qualitative sense of this dependence at present, it seems that
heating rates increase with rf voltage, up to a point, at which
the effect levels off.  This rf-voltage dependence is observed 
along directions where the ion is confined both by static fields and by pondermotive fields.
 It may not be unreasonable that the 
increased rf voltage increases the
intensity of the noise source (possibly due to an increase of temperature of the
electrodes), even when it does not affect the trap secular frequency, as in the 
axial direction of the linear traps.

\begin{figure}
\epsfig{file=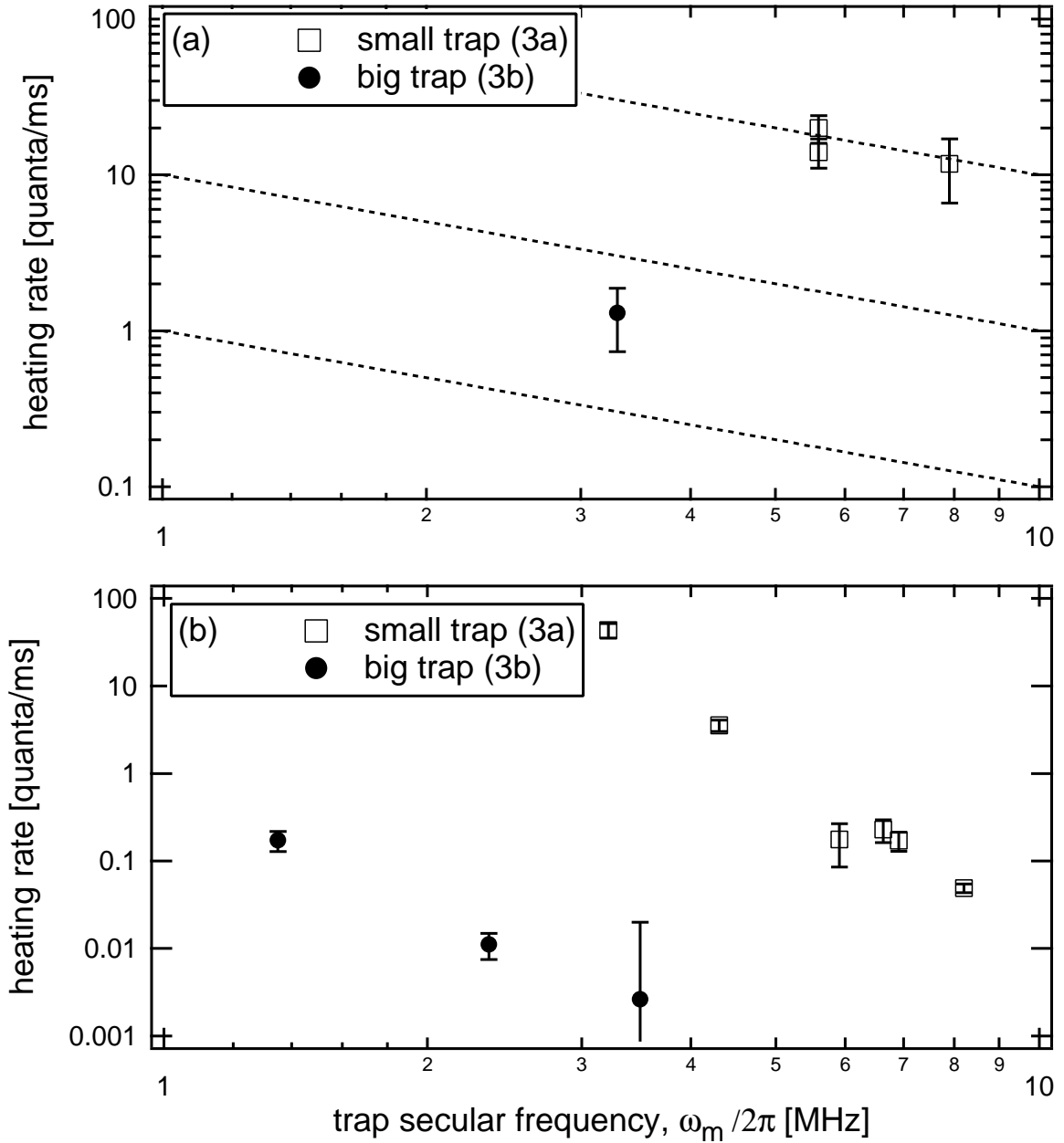,width=\textwidth}
\caption{Data from trap 3, showing heating rates {\it vs.} trap secular frequency.  
(a) Data set \# 1.  The two points on the small
trap at $\omega_m/2\pi = 5.3$ MHz were taken with Raman cooling to $\bar n(t=0) \sim 0$
and with Doppler cooling only to $\bar n(t=0) \sim 2$.  Note that they give
comparable results, as they should.  The dashed lines show a $\omega_m^{-1}$ scaling. 
(b) Data set \# 2.  The small trap data were taken with an rf voltage of $\sim$400 V
and the big trap with  $\sim$600 V. The secular frequency was changed by
tuning the DC bias between fork and ring electrodes. 
}
\label{fig:dtrapd}
\end{figure}

 Trap 3 was designed
to give a controlled measure of the heating rate as a function
of trap size, while all other parameters were held fixed.
The trap electrodes were made from the same substrates,
the electrodes were subjected to the same pre-use cleaning, 
the traps were in the same
vacuum envelope, driven by the same rf electronics (simultaneously) and
data for both traps were acquired with minimal delay.
For direct comparison at the same secular frequency in both traps,
it was necessary to change
the applied rf voltage
since $\omega_m \propto 1/d^2$.
(A static bias between ring and endcap can be added,
 as discussed above, but this was not
sufficient to measure heating at identical secular frequencies
for the same rf drive.)
There are two data sets to be discussed for this trap,
shown in Figure \ref{fig:dtrapd}.

In the first set, shown in Figure \ref{fig:dtrapd}a,
we have data points at two different secular frequencies for the ``small
trap'' (trap 3a) and one point for the ``big trap'' (trap 3b). 
The heating rates of the small 
trap are comparable to the heating rates for other traps and
show a $\omega_m^{-1}$ scaling of the heating rate.  
The single point on the big trap is
at a lower secular frequency, yet has a much 
{\it slower} heating rate.  In fact,
if we extrapolate the data from the small trap to the same secular
frequency (using $\omega_m^{-1}$), 
the heating rate is over an order of magnitude lower in the big trap.
The ratio of the heating rate in the small trap to that of the
big trap is $20 \pm 6$.
This is a much stronger scaling than that predicted by a Johnson noise
heating mechanism (Eq. \ref{eq:heatrateR} predicts a 
$d^{2} \sim 4.8$ scaling),
but is consistent with the scaling in the patch-potential case 
(Eq. \ref{eq:heatrateP} predicts a $d^{4} \sim 23$ scaling).
When these data are used to predict an exponent for the
size-scaling, the result is $d^{3.8 \pm 0.6}$.

For the second data set, shown in Figure \ref{fig:dtrapd}b,  
the trap was removed from the vacuum enclosure, given the
usual cleaning (as discussed above), 
and replaced for the measurements.  In this data set, the 
trap behaved quite differently from all other traps, 
with heating rates significantly below those of other traps.
Also, $S_E$ must have been a strong function of $\omega$ for this trap
since the scaling with trap frequency was 
rather pronounced.  
The scaling with size was also strong: the heating rate
was 16,000 times smaller in the big trap. 
When these data are used to predict an exponent for the
size-scaling, the result is $d^{12 \pm 2}$.
Needless to say,
it is difficult to draw general conclusions from the data for 
this particular trap, but the difference in
heating rates between the two traps seems to strongly indicate, again,
that Johnson noise is not the source of the heating.  We cannot be sure
why this trap had such anomalous heating behaviour, but we speculate that 
it is due to a less-than-usual deposition of Be on the 
trap electrodes prior to the measurements, because the trap loaded
with minimal exposure to the Be source atomic beam. 

At this point it is useful to compare 
the present results to heating rates in other experiments.
There are two other measurements.
The first was done with $^{198}$Hg$^+$ \cite{Diedrich89}.  
For that experiment $\omega_m/2\pi \approx 3$ MHz 
and $d \approx 450 \; \mu$m
and the heating rate was 0.006/ms.
Accounting for scalings with 
trap frequency ($\omega_m^{-1}$) \cite{whylinear} and mass ($m^{-1}$),
these results are consistent with the present results 
for a size-scaling of $d^{-4}$.  
Another measurement has been made with $^{40}$Ca$^+$
\cite{Roos99}. For that experiment $\omega_m/2\pi \approx 4$ MHz
and $d \approx 700 \; \mu$m and the heating rate was 0.005/ms.
Compared to the present experiments and the Hg experiment, this is 
also consistent with a $d^{-4}$ scaling, although it is certainly 
unlikely that all systems had the same patch field environment.

\section{Conclusions and outlook}

We have measured heating from the ground state of trapped ions in
different traps. The magnitude of heating rates and the 
results of the size-scaling measurements are inconsistent with 
thermal electronic noise as the source of the heating.  The results
do not indicate any strong dependence on trap-electrode material
or on the type of trap potential (pondermotive or static).  The
rf voltage applied to the electrodes may play a role in heating, 
in as much as it may have an influence on patch-potentials.

Since we have not identified the mechanism for the observed heating, it is
difficult to say what path should be taken to correct it.  
If fluctuating patch-potentials on the surface 
of the electrodes are the cause, then further cleaning may be appropriate.
Additionally, better masking of the the trap electrodes from the
Be source ovens may help. 

The results coupled with those of other experiments \cite{Diedrich89,Roos99},
strongly indicate that bigger traps have smaller heating rates.
This is not a surprise, but the strength of the scaling may be.  With little
sacrifice in the trap secular frequency (which ultimately determines
the fastest rate of coherent manipulation) a dramatic decrease in the 
heating rate vs. logic gate speed appears possible using larger traps.

We acknowledge support from the U.\ S.\ National Security Agency, Office of 
Naval Research and Army Research Office.
We thank Chris Langer, Pin Chen, and Mike Lombardi for
critical readings of the manuscript.

\section{Appendix: Thermal Electric fields}

We are interested in the thermal electric field power spectral density $S_{E_i}(\bbox{r},\omega)$ generated
from a specified volume of conductor. The conductor can be decomposed into a web of resistors each 
carrying current spectral density $S_{I_i} = 4k_B T/R_i$ (where we assume $k_BT\gg \hbar\omega$). 
The resistance along the $ith$ direction of an infinitesimal volume element is $R_i = dl/(\sigma dA)$, where
$\sigma$ is the conductivity, $dl$ is the length along $i$ and $dA$ is the cross-sectional area.  
A Fourier component of current $I_i(\omega)$ through the volume $dV=dldA$ 
gives rise to an electric dipole $P_i(\omega)=I_i(\omega)dl/\omega$, thus 
the equivalent spectral density of electric dipole of the infinitesimal resistor is isotropic:
$S_{P_i}(\omega) = 4k_BT\sigma dV/\omega^2$.  

The electric field from an electric dipole $\bbox{P}(\bbox{r}',\omega)$ oscillating at frequency 
$\omega$ and position $\bbox{r}'$ is
\begin{equation}
\label{Egreen}
E_i(\bbox{r},\omega)=\sum_{j=x,y,z} P_j(\bbox{r}',\omega)G_{ij}(\bbox {r},\bbox {r}',\omega).
\end{equation}
In this expression, $G_{ij}(\bbox {r},\bbox {r}',\omega)$ 
is a Green function matrix, representing the $ith$ component of electric field at position $\bbox{r}$ 
due to the $jth$ component of a point dipole at $\bbox{r}'$ which satisfies the appropriate boundary 
conditions of the geometry.  The electric field spectral density at position $\bbox{r}$ is 
an integral over the dipoles in the conductor volume:
\begin{equation}
\label{greenint}
S_{E_i}(\bbox{r},\omega) = \frac{4k_BT}{\omega^2}
\int\sigma(\bbox{r}')\sum_{j=x,y,z}\vert G_{ij}(\bbox {r},\bbox{r}',\omega)\vert^2dV'.
\end{equation}

The Green function satisfies $G_{ij}(\bbox {r},\bbox{r}',\omega)=G_{ij}(\bbox {r}',\bbox{r},\omega)$,
so the above integral can be interpreted as the Ohmic power absorbed by the conductor from the
electric fields generated by a point dipole at position $\bbox{r}$.  By energy conservation,
this must be equivalent to the time-averaged power dissipated by a point dipole at $\bbox{r}$, 
which is related to the imaginary part of the Green function matrix 
$G_{ij}(\bbox {r},\bbox{r},\omega)$ \cite{Landau}.  This simplifies
Eq. (\ref{greenint}), leaving the fluctuation-dissipation theorem
\begin{equation}
\label{flucdiss}
S_{E_i}(\bbox{r},\omega)=
\frac{2k_BT}{\omega}\sum_{j=x,y,z}\Im m \: G_{ij}(\bbox {r},\bbox {r},\omega).
\end{equation}

Agarwal solved Maxwell's equations for $G_{ij}(\bbox {r}',\bbox {r},\omega)$ for the simple geometry
of an infinite sheet of conductor filling the space $z\le0$ with the conductor-vacuum interface
in the $z=0$ plane \cite{Agarwal}.  Although this idealized geometry is far from any real ion trap 
electrode structure, rough scalings of the thermal fields can be relevant to real ion trap geometries. 
From Ref. \cite{Agarwal}, the Green function matrix for this problem is diagonal with axial ($z$) and radial ($\rho$) components
\begin{equation}
\label{greenz}
G_{zz}(z,z,\omega) = G^{free}(\omega) + i\int_{0}^{\infty}\frac{q^3}{w_0} 
\left(\frac{w_0\varepsilon-w}{w_0\varepsilon+w}\right)e^{2iw_0z}dq
\end{equation}
\begin{equation}
\label{greenx}
G_{\rho\rho}(z,z,\omega) = G^{free}(\omega)
-\frac{i}{2}\int_{0}^{\infty}q \left[w_0\left(\frac{w_0\varepsilon-w}{w_0\varepsilon+w}\right)
+\frac{k^2}{w_0}\left(\frac{w-w_0}{w+w_0}\right)\right]e^{2iw_0z}dq,
\end{equation}

In the above expressions, $\varepsilon(\omega)=\varepsilon_0+i\sigma/\varepsilon_0\omega$ is the 
dielectric function of the conductor (in the low frequency limit), $k=\omega/c$, 
and wavevectors $w_0$ and $w$ (generally complex) are defined by  
$w_0^2 = k^2-q^2$ and $w^2 = k^2\varepsilon-q^2$ with $\Im m \: w_0\ge0$ and $\Im m \: w\ge0$.
The free space Green's function $G^{free}(\omega)$ has imaginary part 
$\Im m\: G^{free}(\omega)=k^3/6\pi\varepsilon_0$ and gives rise to the isotropic free space blackbody electric 
field fluctuations when substituted into Eq. (\ref{flucdiss}).

The above integrals are significantly simplified in the ``quasi-static" limit, where $kz\ll 1$ and the
conductivity is sufficiently high so that $k\delta\ll 1$, where
$\delta=\sqrt{2c^2\varepsilon_0/\omega\sigma}$ is the skin-depth of the conductor.  Despite
these conditions, no restriction is placed on the value of $z/\delta$.
We break the above integrals into two pieces.  The first piece $\int_0^k$ has $q\le k$ with $w_0$ real.  
In the quasi-static limit, this piece can be shown to cancel the free space contribution 
to the transverse Green function $\Im m\: G_{\rho\rho}(z,z,\omega)$ while doubling the
free space contribution to the axial Green function $\Im m\: G_{zz}(z,z,\omega)$.  Physically, 
the presence of the conductor negates the transverse free space blackbody field while it doubles the axial
blackbody field due to a near-perfect reflection.  The second piece of the integrals $\int_k^{\infty}$ has 
$q \ge k$ with $w_0$ imaginary.  These pieces of the integral can be solved to lowest order in $kz$ and
$k\delta$.  Combining terms and substituting the results into Eq. (\ref{flucdiss}), the thermal electric
field spectral density is 
\begin{equation}
\label{zthermalfield}
S_{E_z}(z,\omega) = \frac{2k_BT\omega^2}{3\pi\varepsilon_0c^3} + 
\frac{k_BT}{4\pi\sigma z^3}
\sqrt{\frac{1}{2}+\sqrt{\frac{1}{4}+\frac{z^4}{\delta^4}}}
\end{equation}
\begin{equation}
\label{rthermalfield}
S_{E_\rho}(z,\omega) = \frac{k_BT}{8\pi\sigma z^3}
\sqrt{\frac{1}{2}+\sqrt{\frac{1}{4}+\frac{z^4}{\delta^4}}}.
\end{equation}

\begin{figure}
\epsfig{file=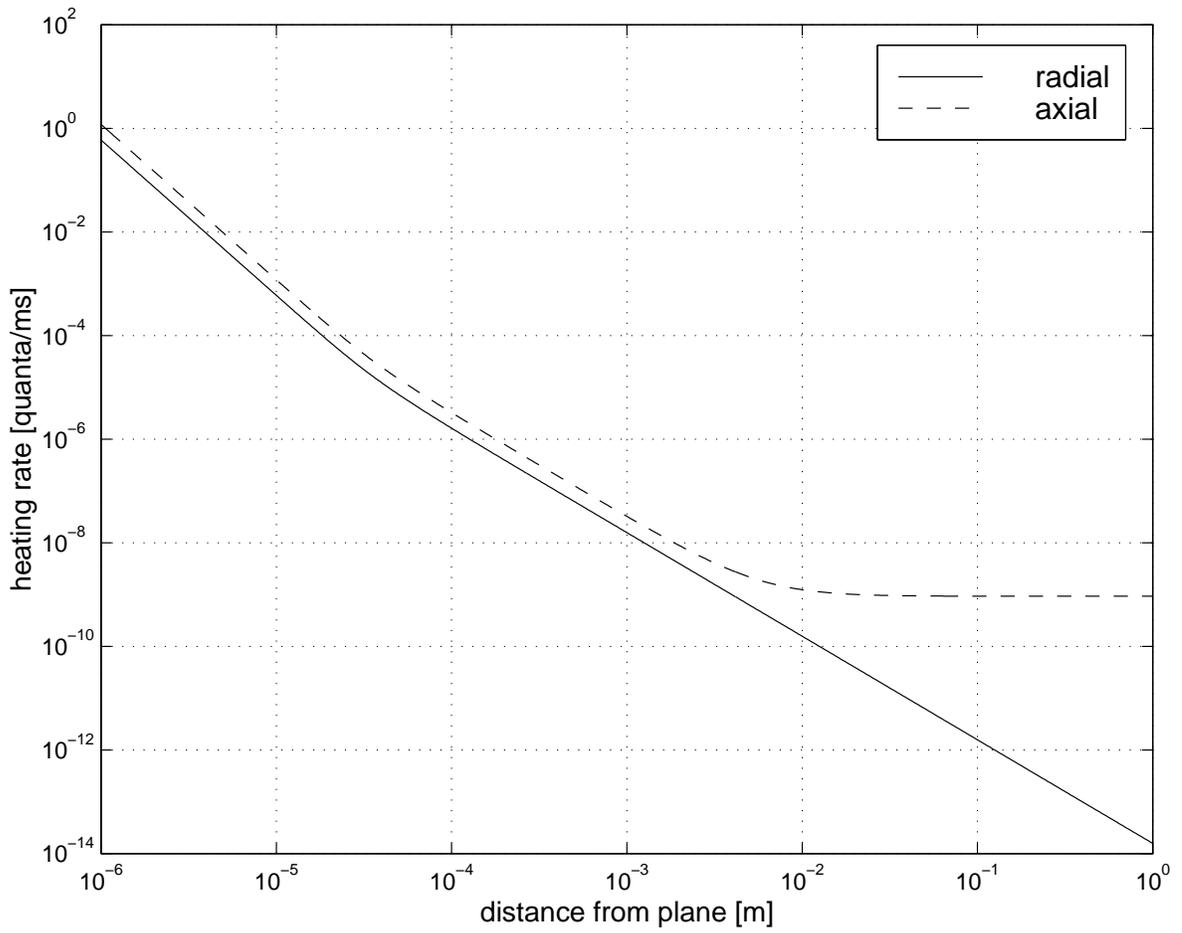,width=\textwidth}
\caption{Expected heating rate {\it vs.} distance from electrode for thermal electronic noise, from
Eqs. (\protect\ref{zthermalfield}), 
(\protect\ref{rthermalfield}) and  (\protect\ref{eq:heatrate}).  The numerical
parameters are those for $^9$Be$^+$ with molybdenum electrodes at 10 MHz secular frequency
and room temperature.
}
\label{fig:therm}
\end{figure}

These expressions show that the thermal electric field noise scales as $1/z^3$ for $z\ll \delta$ \cite{Henkel99}, but scales 
as $1/z^2$ for $z\gg \delta$ \cite{Wineland98,Wineland98a}.  At large distances $z>\sqrt{\delta/k}$ (with $kz \ll 1$), the axial 
field noise settles toward twice the free space blackbody value while the
radial field vanishes. This result is also reported in Ref. \cite{Henkel99a}. 
The behavior is shown in Fig.\ \ref{fig:therm}, where Eqs. (\ref{zthermalfield}) and 
(\ref{rthermalfield}) have been substituted into Eq.\ \ref{eq:heatrate},
 giving the expected thermal heating rate for a $^9$Be ion
trapped with molybdenum electrodes at room temperature.  Note that the predicted heating rate
at trap sizes typical in our experiments is significantly slower than the 0.1-1 quanta/s
rate predicted in \cite{Wineland98,Wineland98a}.  This difference comes from the choice of
the value of the resistance in Eq.\ \ref{eq:heatrateR}, which was chosen in \cite{Wineland98,Wineland98a}
 as an absolute upper limit.  

When interpreting these results, only the rough scaling should be considered.  
Realistic ion trap electrode geometries are more complicated than a single infinite conducting plane, 
involving a more closed electrode structure.  This generally requires a full numerical solution to the 
relevant boundary value problem.  Moreover, we are usually interested in the electric field fluctuations at 
the center of the trap, where these fluctuations will be substantially
different from those above an infinite plate.

\newpage

\newpage

\newpage

\begin{table}
\begin{tabbing}
\hspace{0.5in} \= \hspace{1.1in} \= \hspace{0.7in} \= \hspace{0.75in} \=\hspace{1in} \= \hspace{1in} \= \hspace{0.5in} \\
trap \>type \>material\>size [$\mu$m]\>$\Omega_T/2\pi$ [MHz]\>$\dot{\bar n}$ [ms$^{-1}$]\>ref.  \\
1\>circular ring \> Mo \> 170 \> 250\>1\>\cite{Monroe95a,Meekhof96}   \\
2\>elliptical ring \> Be \> 175 \>250\>10 \> \cite{King98,Turchette98}   \\
3a\>circular ring \> Mo \> 175\>150 \>$10, 10^{-2}$\>    \\
3b\>circular ring \> Mo \> 395\>150 \>$0.5, 10^{-5}$\>    \\
4\>linear \> Au \> 280\>150 \>2.3\> \cite{Myatt99}  \\
5\>linear \> Au \> 280\>230 \>3.5\> \cite{Sackett99}  \\
6\>linear \> Au \> 365\>230 \>1.1\> \cite{Sackett99} \\
\end{tabbing}
\caption{Summary of traps. The size column is approximately
the distance between the ion and the nearest electrode surface.
$\Omega_T$ is the trap rf-drive frequency. The heating rate ($\dot{\bar n}$) is
for a trap secular frequency of 10 MHz, which in the
case of traps 3a and 3b had to be extrapolated from data
at lower trap secular frequencies. The two numbers quoted for 3a and 3b
are for two different versions of the trap.
See the text for a further discussion of the parameters.
}
\label{table:traps}
\end{table}


\begin{references}
\bibitem[*]{gov} Work of the US government.  Not subject to US copyright.
\bibitem{whylinear}  There is no way to know if $\omega_m^{-1}$ is the proper
                     way to scale for trap frequency, but it is a reasonable first assumption.
\end{references}
\end{document}